\documentclass[preprint,12pt]{elsarticle}

\usepackage{amsmath} 
\usepackage{multirow} 
\usepackage{xcolor} 

\usepackage{amssymb}

\usepackage{bm}

\newcounter{bla}

\journal{Computer Physics Communications}

\newcommand{\QPROP}{\textsc{Qprop}}
\newcommand{\IP}{I_{\mathrm{p}}}
\newcommand{\UP}{U_{\mathrm{p}}}
\newcommand{\RI}{R_{\mathrm{I}}}
\newcommand{\imag}{\mathrm{i}}
\newcommand{\MAX}{\mathrm{max}}
\newcommand{\MIN}{\mathrm{min}}
\newcommand{\IM}{\mathrm{im}}
\newcommand{\RCO}{R_{\mathrm{co}}}

\begin{document}

\begin{frontmatter}



\title{\QPROP\ with faster calculation of photoelectron spectra}


\author{Vasily Tulsky}
\author{Dieter Bauer\corref{author}}

\cortext[author] {Corresponding author.\\ \textit{E-mail address:} dieter.bauer@uni-rostock.de (D. Bauer).}
\address{Institut f\"ur Physik, Universit\"at Rostock, 18051 Rostock, Germany}

\begin{abstract}
  The calculation of accurate photoelectron spectra (PES) for
  strong-field laser-atom experiments is a demanding computational
  task, even in single-active-electron approximation. The \QPROP\
  code, published in 2006, has been extended in 2016 in order to
  provide the possibility to calculate PES using the so-called t-SURFF
  approach [L.\ Tao, A.\ Scrinzi, New J.\ Phys.\ 14, 013021
  (2012)]. In t-SURFF, the flux through a surface while the laser is
  on is monitored. Calculating PES from this flux through a surface
  enclosing a relatively small computational grid is much more
  efficient than calculating it from the widely spread wavefunction at
  the end of the laser pulse on a much larger grid. However, the
  smaller the minimum photoelectron energy of interest is, the more
  post-propagation after the actual laser pulse is necessary. This
  drawback of t-SURFF has been overcome by Morales {\em et al.} [F.\
  Morales, T.\ Bredtmann, S.\ Patchkovskii, J.\ Phys.\ B: At.\ Mol.\
  Opt.\ Phys.\ 49, 245001 (2016)] by noticing that the propagation of
  the wavefunction from the end of the laser pulse to infinity can be
  performed very efficiently in a single step. In this work, we
  introduce \QPROP~3.0, in which this single-step post-propagation
  (dubbed i-SURFV) is added. Examples, illustrating the new feature,
  are discussed. A few other improvements, concerning mainly the
  parameter files, are also explained.
\end{abstract}


\end{frontmatter}



{\bf NEW VERSION PROGRAM SUMMARY}

\begin{small}
\noindent
{\em Manuscript Title:} \QPROP~with faster calculation of photoelectron spectra\\
{\em Authors:} Vasily Tulsky, Dieter Bauer\\
{\em Program Title:} Qprop\\
{\em Journal Reference:} \\
{\em Catalogue identifier:}\\
{\em Licensing provisions:} GNU General Public License, version 3\\
\\
{\em Programming language:} C++\\
{\em Computer:} x86\_64\\
{\em Operating system:} Linux\\
{\em RAM:} The memory requirement depends on the desired resolution of the photoelectron spectrum: examples provided with the package need about 1 GB of RAM (with the exception {\tt long-wavelength} that needs 3.5GB of RAM). If no photoelectron spectrum is produced ({\tt hhg} example) then only a few MB are necessary. \\
{\em Number of processors used:} Calculation of PES supports parallelization on up to $N_k$ processes equal to the number of radial momenta of interest.
\\
{\em External routines/libraries: } GNU Scientific Library, Open MPI (optional).\\
\\
{\em Catalogue identifier of previous version:} ADXB\_v2\_0\\
{\em Journal reference of previous version:} Comput. Phys. Comm. 207(2016) 452-463\\
{\em Does the new version supersede the previous version?:} Fully supports the functionality of \QPROP~2.0. \\
{\em Nature of problem:} Efficient calculation of  PES for typical strong-field and attosecond physics ionization scenarios.
   \\
{\em Solution method:} The time-dependent Schr\"odinger equation is solved by propagating the electronic wavefunction using a Crank-Nicolson propagator.
The wavefunction is represented by an expansion in spherical harmonics.
The t-SURFF method in combination with i-SURFV is used to calculate PES.\\
   \\
{\em Reasons for the new version:} The i-SURFV method is employed to speed up the calculation of PES. 
   \\
{\em Summary of revisions:} The i-SURFV method is implemented. A set of examples is provided.\\
   \\
{\em Restrictions:} The atomic potential needs to be of finite range in case of t-SURFF/i-SURFV usage (i.e., the Coulomb tail is truncated at sufficiently large distances). The laser-matter interaction is described in dipole approximation and velocity gauge. \\
   \\
   \\
{\em Additional comments:} For additional information see www.qprop.de\\
   \\
{\em Running time:} Depends on the laser configuration and on the resolution of PES. Most examples require between 7 to 35 minutes. The longest needs about 1.5 hours. \\
   \\
\end{small}

\section{{\bf Introduction}}
Intense-laser-matter experiments brought forward many surprising results that were inaccessible to conventional perturbative theoretical approaches (see, e.g., \cite{Wolter2015}). As a consequence, new but less rigorous or semi-classical methods have been developed  \cite{Kopold2000,Salieres2001,Milosevic2006,Popruzhenko2014a} that, however, need to be tested against numerical {\em ab initio} solutions. Already the solution of the time-dependent Schr\"odinger equation (TDSE) for a single active electron in an effective atomic potential and in the presence of a classical, strong laser field can be a demanding computational task \cite{NumMuller,qprop,Morales_JPhysB_2016,qprop_tsurff}.

The present paper is devoted to a revised version of \QPROP~--- a position-space TDSE-solver for a single active electron bound in a spherically symmetric potential and subject to an external, time-dependent, space-homogeneous electric field (representing the laser field in dipole approximation). \QPROP~was introduced in Ref.~\cite{qprop}. A revised version of \QPROP~employing the time-dependent surface flux method (t-SURFF) for the calculation of PES as proposed in Ref.~\cite{Tao_NewJPhys_2012} was published in Ref.~\cite{qprop_tsurff}.  Using t-SURFF, the PES are calculated from the probability flux through a surface located sufficiently far away from the effective range of the binding potential. The time interval over which the flux is captured is limited by the simulation time. Hence, those components of the electronic wavefunction that represent the slowest electrons of interest should reach the t-SURFF surface during the simulation time. In practice, that means that the simulation time might be many times the actual laser pulse duration, in particular for the simulation of ultra-short pulse experiments. In this paper, we introduce \QPROP~3.0, where this post-pulse propagation just to capture the slow electrons is avoided using the ``trick'' proposed in \cite{Morales_JPhysB_2016} called i-SURFV: once the laser field is off, the evolution of the system is described by a time-independent Hamiltonian, and the contribution to the surface flux after the pulse up to infinity can be calculated in a single step.  Refining the formulas used in \QPROP~2.0 \citep{qprop_tsurff}, it is possible to reduce this evaluation to an action of a non-local operator.

The paper is organized as follows. Section 2 contains the mathematical formulation of the upgraded version of t-SURFF (i.e., i-SURFV) that is implemented in \QPROP~3.0. In Section~3, the most important functions and data structures are described. Section~4 contains examples. Two examples were already in the \QPROP~2.0 paper \cite{qprop_tsurff}, thus demonstrating nicely the improvement in performance using i-SURFV.  A few more demo configurations that may serve as useful templates for a user have been added.

Atomic units $\hbar=|e|=m_e=4\pi\epsilon_0=1$ are used throughout the
paper unless other units are explicitly given.

\section{{\bf Theoretical basis for i-SURFV}}

\subsection{Hamiltonian and wavefunction}
We consider a single active electron, initially bound by the atomic potential, under the influence of a  laser field. This system is described by the TDSE
\begin{equation}\label{schroedinger}
\imag\partial_t\vert\Psi(t)\rangle=\hat{H}(t)\vert\Psi(t)\rangle
\end{equation}
with the Hamiltonian in velocity gauge
\begin{equation}
\hat{H}(t) = -\frac{\Delta}{2}-\imag {\bf A}(t)\cdot\nabla+U(r)-\imag V_{\textrm{\scriptsize{ im}}}(r).
\end{equation}
Here, $U(r)$ is the binding potential of the atom, ${\bf A}(t)$ is the vector potential in dipole approximation (i.e., the electric field is ${\bf E}(t) = -\partial_t {\bf A}(t)$),  and $V_{\textrm{\scriptsize{ im}}}$ is the imaginary potential which plays the role of an absorber to exclude unphysical reflections in the wavefunction  off the numerical boundary. The purely time-dependent  term $\sim {\bf A}^2(t)$ (that arises from the minimum coupling term $(\hat{{\bf p}} + {\bf A}(t))^2/2$)  has been transformed away.  Due to the spherical symmetry of the potential $U(r)$, it is convenient to expand the wavefunction in spherical harmonics,
\begin{equation}
\left<{\bf r}\vert \Psi(t)\right> = \frac{1}{r}\sum_{\ell=0}^{\infty}\sum_{m=-\ell}^{\ell}\phi_{\ell m}(r,t)Y_{\ell m}(\Omega).
\end{equation}
The upper limit for the orbital angular momentum quantum number $\ell$ is finite in numerical calculations, i.e.,
\begin{equation}\label{psi1}
\langle {\bf r}\vert \Psi(t)\rangle = \frac{1}{r}\sum_{\ell=0}^{N_\ell-1}\sum_{m=-\ell}^{\ell}\phi_{\ell m}(r,t)Y_{\ell m}(\Omega).
\end{equation}
$N_\ell$ is defined by the user.  If the polarization of the laser is chosen linear along the $z$ axis and the magnetic quantum number $m_0$ of the initial state is well defined, $m_0$  is conserved during time propagation so that only $m_0$ contributes to the sum over $m$,
\begin{equation}\label{psi2}
\langle {\bf r}\vert \Psi_{m_0}(t)\rangle = \frac{1}{r}\sum_{\ell=\vert m_0 \vert}^{N_\ell-1}\phi_{\ell m_0}(r,t)Y_{\ell m_0}(\Omega).
\end{equation}
We discretize time and the radial coordinate in units of $\Delta t$ and $\Delta r$, respectively. The radial grid is of size $R_{\MAX}=\RI+\alpha+W_{\IM}$ where $\RI$ is the position of the flux-capturing surface for t-SURFF, $\alpha=E_{\MAX}/\omega^2$ is the quiver amplitude in the chosen laser field of electric field maximum $E_{\MAX}$, and $W_{\IM}$ is the width of the imaginary potential that has the form
\begin{equation}
V_{\IM}= V_{\IM,\MAX} \left(\frac{r-R_{\IM}}{W_{\IM}}\right)^{\!\!16}\Theta(r-R_{\IM}),
\end{equation}
with $R_{\IM}=\RI+\alpha$, and $V_{\IM,\MAX}=100$ by default. The spherically symmetric potential $U(r)$ can be defined by the user.  In the following, we use  a potential of the residual atom that is hydrogenic but is switched to linear at $r>\RCO$ and is off after reaching zero at $r=2\RCO$:
\begin{equation}
U(r) =
    \begin{cases}
      -1/r & \text{ if } r<\RCO\\
      -(2\RCO-r)/\RCO^2 &\text{ if } \RCO<r<2\RCO\\
      0 &  \text{ if } r>2\RCO
    \end{cases}  .
\end{equation}
The cutoff radius $\RCO$ is defined by the user in the {\tt initial.param} file. The t-SURFF/i-SURFV method for the calculation of PES requires that $\RI > 2\RCO$. The shapes of the binding and imaginary potentials are defined in {\tt potentials.hh}.

\subsection{t-SURFF and i-SURFV}
Before we move on to the i-SURFV method, let us briefly review  the basics of t-SURFF \cite{Tao_NewJPhys_2012} (see \cite{qprop_tsurff} for t-SURFF in the context of \QPROP). The PES amplitudes $a_{\textmd{\tiny{I}}}({\bf k},T)$ at time $T$ after the laser pulse are approximated in t-SURFF  by projecting the part of the wavefunction that is farther away from the origin than $\RI$ onto Volkov states of the momentum of interest ${\bf k}$,
\begin{equation}\label{Volkov}
a_{\textmd{\tiny{I}}}({\bf k},T) = \left<{\bf k}(T)\vert \Theta(r-\RI)\vert\Psi(T)\right>. 
\end{equation}
Volkov states are plane-wave states in the presence of a laser field \cite{Volkov1935,Milosevic2006}.  $\RI$ is chosen such that the potential and, hence, the bound states in the region beyond are irrelevant for the effect studied. The time $T$ should be large enough such that  an electron with the smallest momentum of interest  $k_{\MIN}$ arrives at the t-SURFF boundary $\RI$, i.e., $T\geq \RI/k_{\MIN}$.  The amplitude \eqref{Volkov} can be rewritten in the form of a time integral
\begin{equation}
a_{\textmd{\tiny{I}}}({\bf k},T) = \int_0^T dt ~\partial_t \langle{\bf k}(t)\vert \Theta(r-\RI)\vert\Psi(t)\rangle + \langle{\bf k}(0)\vert \Theta(r-\RI)\vert\Psi(0)\rangle.
\end{equation}
The second term vanishes, as with a properly chosen $\RI$ the initial wavefunction is negligible for $r>\RI$. 
Using the TDSE (\ref{schroedinger}) for $\vert\Psi(t)\rangle$ and $-\imag\partial_t\langle {\bf k}(t)\vert= \hat{H}_0 \langle{\bf k}(t)\vert$ for $\langle{\bf k}(t)\vert$, where $\hat{H}_0 = -\Delta/2-\imag{\bf A}(t)\cdot\nabla$ coincides with $\hat{H}$ in the region $r>\RI$,
the remaining term can be written as
\begin{equation}
a_{\textmd{\tiny{I}}}({\bf k},T) = \imag\int_0^T dt ~\langle{\bf k}(t)\vert \left[ \hat{H},\Theta(r-\RI)\right]\vert\Psi(t)\rangle .
\end{equation}
At this point, let us split the integral over time into parts before and after the end of the laser pulse $\tau_p$,
\begin{align}\label{split}
a_{\textmd{\tiny{I}}}({\bf k},T) &= a_{\textmd{\tiny{I}}}({\bf k},\tau_p) + \delta a_{\textmd{\tiny{I}}}({\bf k},\tau_p,T),\\
\delta a_{\textmd{\tiny{I}}}({\bf k},\tau_p,T) &= \imag\int_{\tau_p}^T dt ~\langle{\bf k}(t)\vert \left[ -\frac{\Delta}{2},\Theta(r-\RI)\right]\vert\Psi(t)\rangle. \nonumber
\end{align}
The first term $a_{\textmd{\tiny{I}}}({\bf k},\tau_p)$ is treated in \QPROP~numerically as it was described in \cite{qprop_tsurff} while the second, field-free term  $\delta a_{\textmd{\tiny{I}}}({\bf k},\tau_p,T)$ can be calculated without expensive post-propagation of the wavefunction after the laser pulse. This avoidance of post-propagation up to the time where the slowest electrons of interest passed the t-SURFF boundary is the essence of i-SURFV as proposed in Ref.~\cite{Morales_JPhysB_2016} and the core advantage of the new version of \QPROP~3.0 over version 2.0. 

Let us rewrite the field-free part in the matrix element (\ref{split}) as
\begin{align}
\delta a_{\textmd{\tiny{I}}}({\bf k},\tau_p,T) = -\imag\int_{\tau_p}^T dt ~\int d\Omega~\int dr r^2~ \Big( \delta(r-\RI) \psi^*_{\bf k}({\bf r},t) \partial_r \Psi({\bf r},t)  \nonumber \\
+ \psi^*_{\bf k}({\bf r},t) \Psi({\bf r},t)\frac{\Delta}{2}\Theta(r-\RI) \Big)
\end{align}
with $\psi^*_{\bf k}({\bf r},t) = \langle {\bf k}(t)\vert {\bf r}\rangle$ and $\Psi({\bf r},t) = \langle {\bf r}\vert \Psi(t)\rangle$. Integrating the last term by parts, we obtain
\begin{equation}\label{a1}
    \delta a_{\textmd{\tiny{I}}}({\bf k},\tau_p,T) = -\frac{\imag \RI^2}{2}\int_{\tau_p}^T dt ~\int d\Omega~\Big( \partial_r \psi^*_{\bf k}({\bf r},t)\Psi({\bf r},t)-\psi^*_{\bf k}({\bf r},t)\partial_r \Psi({\bf r},t)\Big) \Big\vert_{r=\RI}.
\end{equation}
The (complex conjugated) plane-wave final-momentum state in position space at times $t \geq \tau_p$
can be expressed as
\begin{equation}
\langle {\bf k}(t)\vert {\bf r}\rangle = \psi^*_{\bf k}({\bf r},t) =\sqrt{\frac{2	}{\pi}}e^{\imag k^2t/2}\sum_{\ell m}(-\imag )^{\ell} j_{\ell}(kr)Y_{\ell m}^*(\Omega)Y_{\ell m}(\Omega_k)
\end{equation}
where $j_\ell(kr)$ are spherical Bessel functions, 
and $\Psi({\bf r},t)$ is represented as (\ref{psi1}) or (\ref{psi2}). Expanding
\begin{equation}
\delta a_{\textmd{\tiny{I}}}({\bf k},\tau_p,T) = \sum_{\ell m} \delta a_{\textmd{\tiny{I}},\ell m}(k,\tau_p,T)Y_{\ell m}(\Omega_k),
\end{equation}
one finds
\begin{eqnarray}
\lefteqn{\delta a_{\textmd{\tiny{I}},\ell m}(k,\tau_p,T) }\\ \nonumber
&= & \frac{(-\imag )^{\ell+1}}{\sqrt{2\pi}} \bigg\{ \big[ k j'_{\ell}(kr)-j_{\ell}(kr)(1 - \RI\partial_r) \big] \int_{\tau_p}^{T} dt~ e^{\imag k^2t/2} \phi_{\ell m}(r,t)\bigg\} \bigg\vert_{r=\RI}
\end{eqnarray}
with $kj'_{\ell} (kr) = \partial_r j_{\ell}(kr)$. The formula $xj'_{\ell} (x) = -xj_{\ell+1}(x)+\ell j_{\ell}(x)$ is used in the program code to calculate the derivatives of the spherical Bessel functions.


The time integral can be evaluated in the limit $T \rightarrow \infty$ using the explicit form of the time evolution operator,
\begin{equation}\label{greens_func}
\int_{\tau_p}^{\infty} dt~ e^{\imag E t}\phi_{\ell m}(r,t) = \int_{\tau_p}^{\infty} dt~ e^{\imag E t-\imag \hat{H}_{\ell}(t-\tau_p)}\phi_{\ell m}(r,\tau_p) = \frac{\imag e^{\imag E \tau_p}}{E-\hat{H}_{\ell}}\phi_{\ell m}(r,\tau_p)\bigr|_{\RI},
\end{equation}
where the time-independent Hamiltonian has the form
\begin{equation}
\hat{H}_{\ell} = -\frac{\partial_r^2}{2}+\frac{\ell(\ell+1)}{2r^2}+U(r)-\imag V_{\textrm{\scriptsize{im}}}(r).
\end{equation}
The operator $\hat{G}_{\ell}(E)= (\hat{H}_{\ell}-E)^{-1}$ is the Green's function for the radial Schr\"odinger equation, i.e., the energy representation of the solution to $$(\hat{H}'_{\ell}-\imag\partial_t)G_{\ell}(r,t)= \delta(t)\delta(r-r').$$ The related function in the i-SURFV code is thus named\ {\tt gfunc}.
The contribution of the upper limit in (\ref{greens_func}) vanishes due to the presence of an absorbing, imaginary potential. The final expression for the field-free part of the full amplitude can be thus written as
\begin{eqnarray}\label{a1_final}
\lefteqn{\delta a_{\textmd{\tiny{I}},\ell m}(k,\tau_p,T)} \\  \nonumber
& =&\frac{(-\imag )^{\ell}}{\sqrt{2\pi}} \bigg\{ \big[ k j'_{\ell}(kr)-j_{\ell}(kr)(1 - \RI\partial_r) \big] 
\frac{e^{\imag E_k \tau_p}}{E_k-\hat{H}_{\ell}}\phi_{\ell m}(r,\tau_p)\bigg\} \bigg\vert_{r=\RI}
\end{eqnarray}
where $E_k=k^2/2$.
The integral over an infinite time has been converted to a single application of the operator $\left[E_k-\hat{H}_{\ell}\right]^{-1}$ (per $k$ and $\ell$), which is the essence of the infinite-time version of t-SURFF, i.e.,  i-SURFV. How the application of such an inverse operator is actually implemented needs not to be detailed here, as it is similar to the application of the spectral window operator explained in detail in \cite{qprop}.   The applicability of the i-SURFV ``trick'' is discussed in section~\ref{sec:examples} by several example configurations included in the \QPROP~3.0 distribution.

\section{{\bf \QPROP~general structure}}

\subsection*{Parameters and flags}
In the current version, all parameters defining coordinate, momentum, and time grids, the potentials, and the laser are moved to the {\tt *param} files. Most of the flags that allow to switch between different methods or to turn on and off the generation or the storage of specific output are also put into those files. Thus, it is no longer necessary to touch {\tt *.cc} files for a wide range of problems.   All parameters and flags are commented so that their function should become very clear while going through the examples  in section~\ref{sec:examples}.

\subsection*{Functions and classes}

The core parts of the current \QPROP~version are as follows.
\begin{itemize}
\item The real or imaginary time propagation by a single timestep $\Psi(t) \rightarrow \Psi(t+\Delta t)$ is performed in the member function {\tt propagate} of class {\tt wavefunction}, which is described in the first \QPROP\ paper \cite{qprop}.
\item The class {\tt tsurffSaveWF} was designed for saving $\phi_{\ell m}(\RI,t)$ and $\partial_r\phi_{\ell m}(r,t)\vert_{r=\RI}$ required for t-SURFF.  For i-SURFV, an additional class {\tt tsurffSave\_full\_WF} saving $\phi_{\ell m}(r,t)$ was added. The data files end with {\tt *.raw}.
\item The Green's function (\ref{greens_func}) is applied in {\tt gfunc}, calculating $(\hat{H}_{\ell}-E)^{-1}\phi_{\ell m}(r,t)\vert_{r=\RI}$ and $\partial_r(\hat{H}_{\ell}-E)^{-1}\phi_{\ell m}(r,t)\vert_{r=\RI}$ from $\phi_{\ell m}(r,t)$ for given $E$.
\item The calculation of a PES using t-SURFF and i-SURFV is performed with the help of the class {\tt tsurffSpectrum}. 
\end{itemize}

\subsection*{Surface flux output format}
In the \QPROP\ 2.0 paper \cite{qprop_tsurff}, two possible expansions of the amplitudes $a_{\textmd{\tiny{I}}}({\bf k})$ were introduced.
If the expansion in angles (according eq.~(21) in \cite{qprop_tsurff}) is chosen, the data generated by {\tt tsurffSpectrum} ((29) in \cite{qprop_tsurff}) in the output files {\tt tsurff\_polar}$i_p${\tt .dat}\footnote{Here, $i_p$ refers to the index of the process that generated the file ($i_p=0$ if the non-parallelized version is used).} is formatted as

\begin{center}
\begin{tabular}{|l|l|l|l|l|l|l|}
\hline
$E_k$ & $k$ & $\theta$ & $\varphi$ & $k\vert a({\bf k})\vert^2$ & $\textrm{Re}~ a({\bf k})$ & $\textrm{Im}~ a({\bf k})$ \\
\hline
\end{tabular}\ .
\end{center}

For a linearly polarized laser pulse along the $z$ axis, the spectrum does not depend on the azimuthal angle $\varphi$, and the calculation is performed for $\varphi=0$ only. For a laser pulse polarized in the $xy$ plane, the user might be solely interested in the spectrum in the polarization plane. For that purpose, only data for $\theta=\pi/2$ is generated if $N_{\theta}=1$ is chosen.

If a ``complete'' expansion in spherical harmonics ((30) in \cite{qprop_tsurff}) is desired, data with partial amplitudes according eq.~(35) in \cite{qprop_tsurff} are stored in {\tt tsurff\_partial}$i_p${\tt .dat} files in the format
\begin{center}
\begin{tabular}{|l|l|l|l|l|l|l|l|}
\hline
$E_k$ & $k$ & $\textrm{Re}~ a_{0}(k)$ & $\textrm{Im}~ a_{0}(k)$ & $\dots$ & $\textrm{Re}~ a_{N_{\tilde\ell}}(k)$ & $\textrm{Im}~ a_{N_{\tilde\ell}}(k)$ & $\vert\sum a_{\tilde{\ell}}(k)\vert^2$ \\
\hline
\end{tabular} 
\end{center}
where $\tilde{\ell}=\ell$ with $N_{\tilde\ell}=N_\ell-1$ in the case of expansion (\ref{psi2}), and $\tilde{\ell}=(\ell+1)\ell+m$ with $N_{\tilde\ell}=N_\ell^2-1$ in the general case (\ref{psi1}). 

One should note that the complete expansion involves the calculation of Clebsch-Gordan coefficients in terms of  Wigner 3$j$ symbols, which are evaluated using the GNU Scientific Library (GSL). However, the respective GSL routine {\tt gsl\_sf\_coupling\_3j}  appears to have acceptable precision only for relatively small $\ell \lesssim 50$ ( see, e.g., Fig.~1 in \cite{Johansson_SIAM_2016}). Nevertheless, for small $\ell$ the complete expansion can be used to identify dominating angular momenta as a function of energy.

\subsection*{Complexity scaling}
The main steps that are required for the calculation of PES with \QPROP~scale as follows with respect to the number of timesteps, radial grid points, orbital momenta, photoelectron energies and angles:
\begin{itemize}
\item Imaginary-time propagation: $N_{R} N_{t}$
\item Real-time propagation:      $N_{\tilde{\ell}} N_{R} N_{t}$
\item t-SURFF or i-SURFV:         $N_{\tilde{\ell}} N_{t} N_{E} N_{\theta} N_{\varphi}$
\item Green's function ``trick'' for i-SURFV:     $N_{\tilde{\ell}} N_{R} N_{E}$
\end{itemize}

\section{{\bf Examples}} \label{sec:examples}
The quickest way to run an example is to go to its folder and launch the {\tt do\_all} bash script by typing {\tt ./do\_all.sh} in the terminal. Alternatively, one may make and execute the programs for imaginary time propagation ({\tt imag\_prop}), real-time propagation ({\tt real\_prop}) and t-SURFF ({\tt tsurff} or {\tt tsurff\_mpi} for an MPI-parallelized version) manually. These programs have been slightly revised and renamed in \QPROP~3.0.  Table~1 shows the old program names in \QPROP~2.0, the new names in \QPROP~3.0, together with the program task. The user may edit the {\tt tsurff.param} file and set the variable {\tt tsurff-version} equal to 1 for t-SURFF or equal to 2 for i-SURFV. If i-SURFV is chosen, it is required to make and execute {\tt isurfv} after the real-time propagation (this step is included in {\tt ./do\_all.sh}). This  generates the data used to calculate the laser-free contribution to the PES according to eq.~(\ref{a1_final}). Already in \QPROP~2.0, half a Hanning window 
\begin{equation}
h(t)=\sin^2(\pi t/T)
\end{equation}
was by default multiplied to the integrand for the $T/2<t<T$ interval of the time integral (\ref{a1}). We keep this feature in \QPROP~3.0 if t-SURFF is used.

\begin{table}[!h]
\caption{Programs names in \QPROP~2.0 and 3.0.}
\begin{tabular}{|l|l|l|}
\hline 
\QPROP~ 2.0               & \QPROP~ 3.0                & Task                                            \\ \hline
{\tt hydrogen\_im}     & {\tt imag\_prop}        & Imaginary-time propagation                          \\ \hline
{\tt hydrogen\_re}     & {\tt real\_prop}        & Real-time propagation                               \\ \hline
~                         & {\tt isurfv}      & $(\hat{H}_{\ell}-E)^{-1}$ to $\phi_{\ell m}$, $\partial_r \phi_{\ell m}$              \\ \hline
{\tt eval-tsurff}({\tt -mpi}) & {\tt tsurff}({\tt \_mpi})  & Spectrum with t-SURFF \\ \hline
\end{tabular}
\end{table}

All parameters defining the $r\ell$-grid, the laser pulse and the momentum grid for the PES are defined in the {\tt *.param} files. Their precise meaning can be found in the comments there.

All plots shown in this paper were produced using Python scripts that are provided with \QPROP~3.0. A brief guide to them can be found at the end of the paper in section~\ref{sec:plotguide}.

\subsection{Speeding up a previous \QPROP~2.0 example}
{\bf Time\footnote{For this and other examples, the estimated  run times are calculated according [{\tt imag\_prop} + {\tt real\_prop} + {\tt isurfv} time] + ([{\tt tsurff\_mpi} average time] $\times$ 4 processes used) if the i-SURFV method is used. The compilation time is not counted. Calculations are performed on a desktop PC with Intel(R) Core(TM) i5-6500 CPU.}:} 6 min + (1 min $\times$ 4).

{\bf Memory\footnote{The maximum RAM space required is determined by the surface flux that needs to be loaded to calculate the PES with t-SURFF. For the default parameters it is 16 bytes $\times$ 4 (or 3 in the linear-polarization case) $\times N_{\tilde{\ell}} N_k N_{\theta} N_{\varphi}$.}:} 229 MB $\times$ 4.

\begin{figure}
\begin{center}
\includegraphics[scale=0.09]{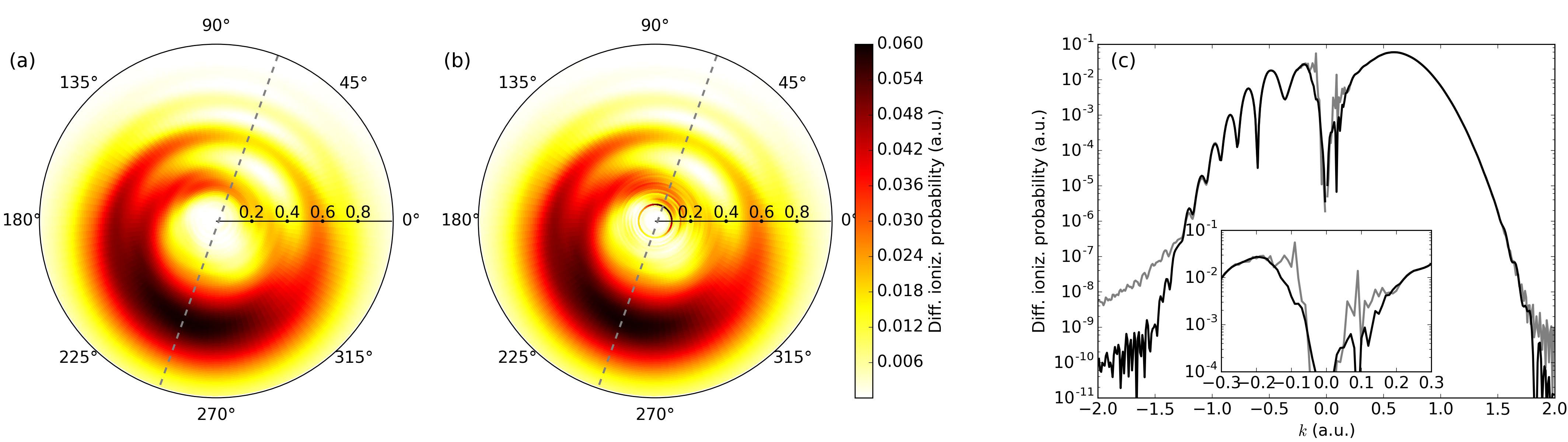}
\end{center}
\caption{PES for hydrogen in a circularly polarized laser ($\lambda=400$\,nm, $I=2 \cdot 10^{14}$ Wcm$^{-2}$, $n_c=2$ cycles, $\sin^2$ envelope) calculated with (a) t-SURFF and (b) i-SURFV. Ticks on the horizontal black line indicate radial momenta. The black and gray spectra shown in (c) and its inset are taken along the dashed lines in (a) and (b), respectively.} \label{fig1}
\end{figure}

In the previous \QPROP ~version, t-SURFF required lengthy post-propagation in real time after the laser pulse is over in order to obtain the yield at low momenta. The time interval necessary to capture the PES down to momentum $k_{\MIN}$ was estimated as $t_{\MIN}=\RI/k_{\MIN}$. In contrast, the i-SURFV real-time propagation up to the end of the pulse plus one additional step covering  $t \to \infty$ is independent of the lowest momentum of interest. The difference between the total calculation times for the ``old'' t-SURFF and the new i-SURFV is especially pronounced for short pulses, which are, in fact, often used in modern experiments. To demonstrate this, we revisit an example from \QPROP~2.0 \cite{qprop_tsurff}: a hydrogen atom in a very short ($n_c=2$ cycles) circularly polarized pulse\footnote{Note that in the new version of \QPROP\ linear or circular laser polarizations are chosen in  {\tt initial.param} where the value of {\tt qprop-dim} is set to $34$ for linear polarization (along $z$ axis) or $44$ for circular polarization (in $xy$ plane). No {\tt *.cc} source files need to be modified anymore.}  with $\omega=0.114$ ($\lambda = 400$\,nm), $E_0=0.0534$ ($\textrm{I}=2 \cdot 10^{14}$ Wcm$^{-2}$) described by the vector potential ${\bf A}(t) = (A_x(t),A_y(t),0)$ (the shape of which is defined in {\tt potentials.hh}),
\begin{eqnarray}
A_x(t) = \frac{E_0}{\omega} \sin^2\left(\frac{\omega t}{2 n_{c}}\right) \sin(\omega t), \\ \nonumber
A_y(t) = \frac{E_0}{\omega} \sin^2\left(\frac{\omega t}{2 n_{c}}\right) \cos(\omega t).
\end{eqnarray}
The corresponding folder is named {\tt attoclock}, eponymous to the
experiments with such pulses \cite{Eckle2008}.  Compare Fig.~\ref{fig1}(a) with Fig.~\ref{fig1}(b) that
represent the results obtained with t-SURFF and i-SURFV,
respectively. The real-time propagation for i-SURFV is
equal to the pulse duration and takes 2205 timesteps. It is significantly smaller than the total time required for
t-SURFF with $k_{\MIN}=0.1$, which takes 22205 timesteps. 
Fig.~\ref{fig1}(c) shows the spectra cut
along $250^\circ$ (and $70^\circ$, represented by negative $k$ values). A
discrepancy on the logarithmic scale is visible in the high-energy
region close to the noise level. This could be improved by increasing the values for  $\RI$
and $W_{\IM}$, i.e.,  {\tt R-tsurff} and {\tt imag-width} in {\tt propagate.param}. 
The low-energy part of the i-SURFV spectrum has a ring-shaped, sharp
peak. This artifact is not caused by i-SURFV
and would appear in t-SURFF as well if a longer laser-free post-propagation
were performed (i.e., if a smaller $k_{\MIN}$ were chosen). The
origin of this structure is the truncation of the potential, and it can be
shifted to even lower energies by increasing the cutoff radius 
$\RCO$ (i.e., {\tt pot-cutoff} in {\tt initial.param}) or by choosing a smoother  potential shape in the intermediate domain $\RCO < r < 2
\RCO$. As noted above, the potential should vanish at
the flux-capturing surface and beyond so that  Volkov states can be used in (\ref{Volkov}).

Besides the artificial sharp peak, an unphysical oscillatory behavior
is present in the low-energy region of Fig.~\ref{fig1}(b). This was
also observed in \citep{Morales_JPhysB_2016} and, in fact, is
not due  to the i-SURFV trick itself but because of  the
imaginary potential. Upon increasing $W_{\IM}$ for the final i-SURFV step according (\ref{greens_func}), those rings go to
lower energies and decrease in magnitude. For that reason  a parameter {\tt
  isurfv-imag-width-factor} is added to the {\tt tsurff.param}
file. 

\begin{figure}
\begin{center}
\includegraphics[scale=0.08]{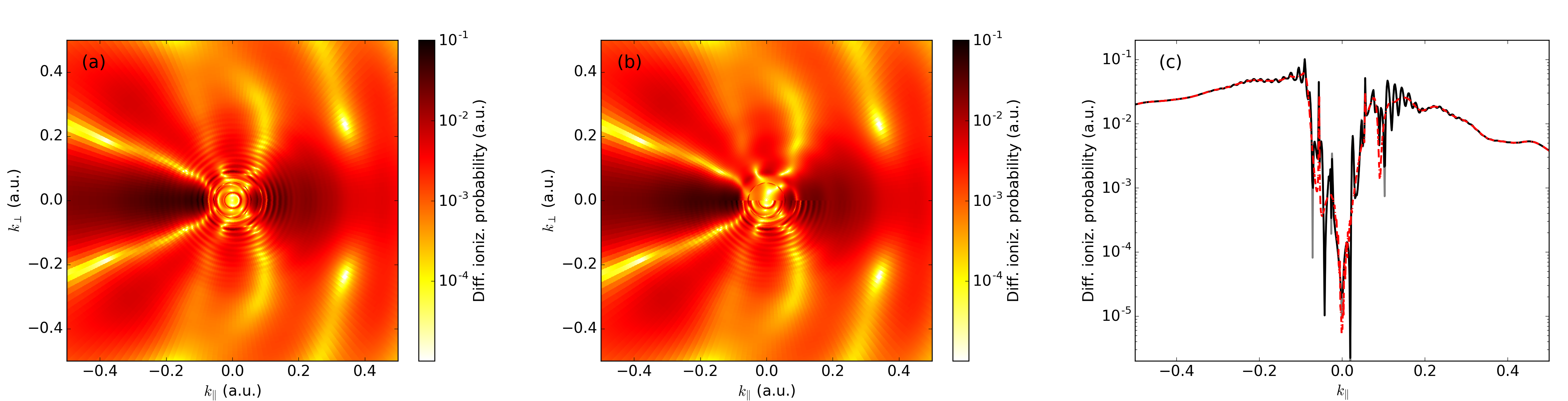}
\end{center}
\caption{PES for hydrogen in a linearly polarized pulse ($\lambda=400$\,nm, $I= 10^{14}$ Wcm$^{-2}$, $n_c=2$ cycles, $\sin^2$ envelope). (a) t-SURFF with $k_{\MIN}=0.001$ (upper halfplane) and i-SURFV (lower halfplane), $W_{\IM}=150$ for both. (b) i-SURFV with $W_{\IM}=1500$ (upper halfplane) and i-SURFV with $W_{\IM}=150$ (lower halfplane). (c) Cut along $k_\parallel$ at $k_{\perp}=0$ for t-SURFF (black solid), i-SURFV with $W_{\IM}=150$ (gray solid) and i-SURFV with $W_{\IM}=1500$ (red dashed-dotted).} \label{fig2}
\end{figure}

Figure~\ref{fig2} illustrates the disappearance of these oscillations in the case of linear polarization with the pulse defined according (\ref{Az}). Further, $k_{\MIN}=0.001$, $N_k=2000$,
$N_{\theta}=101$, $N_{\varphi}=1$ were set in {\tt tsurff.param}. Note that
only the sharp peak around $k=0.056$ caused by the truncation of the
potential survives, and the oscillations due to the imaginary potential are shifted to 10 times smaller
momentum.

Alternative ways to deal with the absorber-produced artifacts exist. For instance, a special shape of $V_{\IM}(r)$ to supress reflections and transitions might be chosen (see, e.g., \cite{Manolopoulos_JChemPhys_2002}) and defined accordingly in {\tt potentials.hh}. However, this approach is energy-dependent. Another way is to employ complex scaling in the absorption region instead of an imaginary potential \cite{Moiseyev_JPhysB_1998, Scrinzi_PhysRevA_2010}. Complex scaling is not yet implemented in \QPROP\ though.

\subsection*{Long wavelength}

{\bf Time:} 46 min + (54 min $\times$ 4).

{\bf Memory:} 879 MB $\times$ 4.

\begin{figure}
\begin{center}
\includegraphics[scale=0.12]{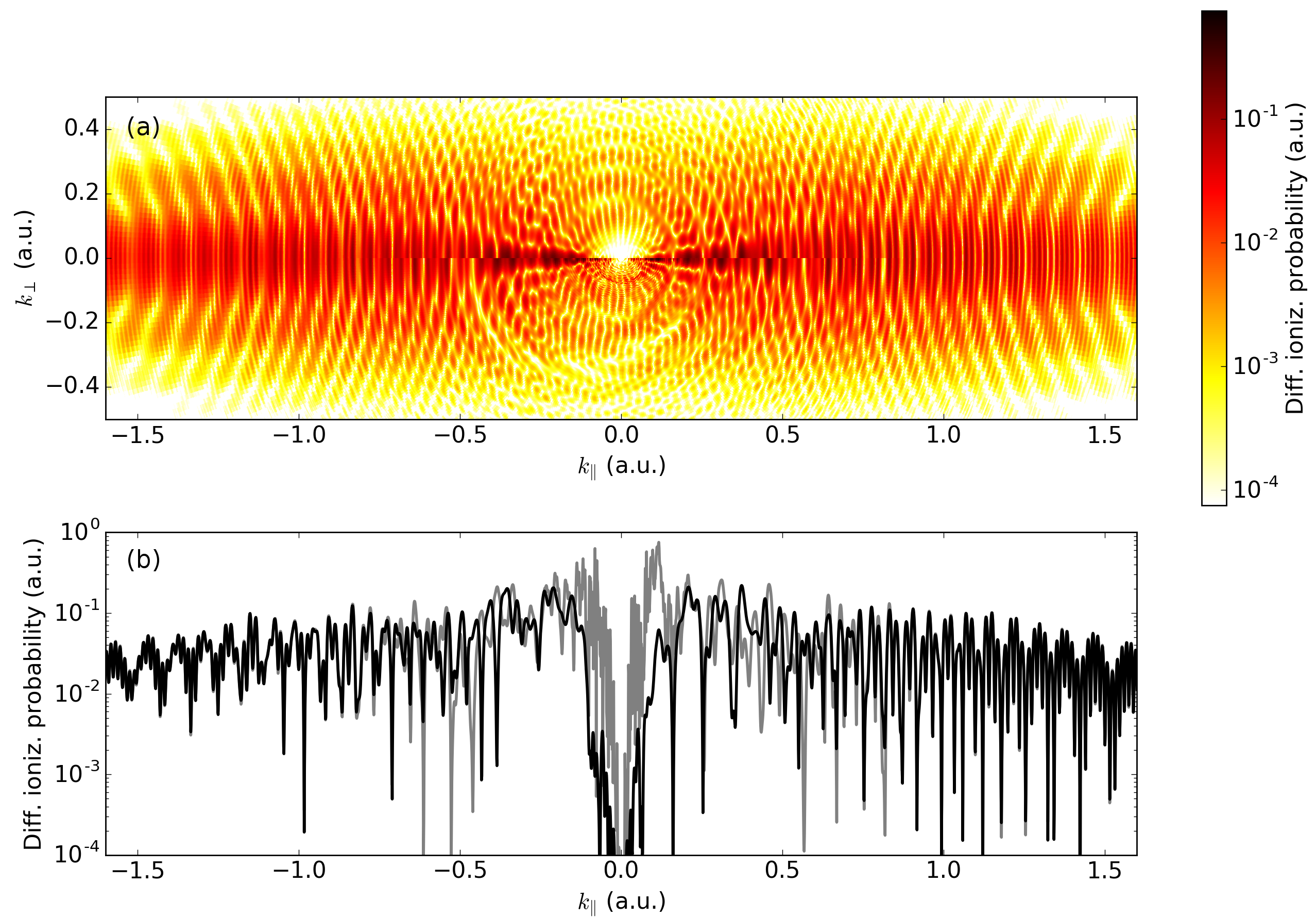}
\end{center}
\caption{(a) PES for hydrogen in a linearly polarized laser pulse ($\lambda=2000$\,nm, $I=10^{14}$ Wcm$^{-2}$, $n_c=6$ cycles, $\sin^2$ envelope) calculated with t-SURFF (upper halfplane) and i-SURFV (lower halfplane).  (b) PES along $k_\parallel$ at $k_{\perp} = 0$, black for t-SURFF, gray for i-SURFV.} \label{fig3}
\end{figure}

We now reconsider the most resource-greedy example in
Ref.~\cite{qprop_tsurff}: the calculation of the PES for atomic hydrogen
subject to a 6-cycles laser pulse
\begin{equation}\label{Az}
A_z(t) = \frac{E_0}{\omega} \sin^2\left(\frac{\omega t}{2 n_{c}}\right) \sin(\omega t)
\end{equation}
with long wavelength $\lambda=2000$\,nm and intensity $I=10^{14}$
Wcm$^{-2}$. The sources are located in the folder {\tt
  long-wavelength}.  In Ref.~\cite{qprop_tsurff}, the focus was on the
$10\UP$ rescattering regime ($\UP = E_0^2/4\omega^2$ is the
ponderomotive energy).  The low-energy part of the PES was not of
interest so that post-propagation after the laser pulse was not an
issue. Now we apply i-SURFV so that we can examine the low-energy part
for this setup, completing the description of the entire spectrum. To
speed calculations up, the upper limit $N_\ell$ for the angular
momenta is decreased to $150$, which is possible because we do not
need to describe the high-energy electrons with high accuracy
anymore. In fact, $N_\ell = 150 > (\IP + 2 \UP)/\omega$ was chosen instead of
$(\IP+10 \UP)/\omega=623$ required in
Ref.~\cite{qprop_tsurff}. Figure~\ref{fig3} shows spectra obtained
with t-SURFF and i-SURFV. Here, $k_{\MIN}$ for t-SURFF was chosen 0.1
instead of 0.5 in \cite{qprop_tsurff}. This slows down the ``old''
t-SURFF approach because of the longer post-propagation time (93070
timesteps in total).  The photoelectron yield near zero momentum is
still not converged. Instead, i-SURFV requires just 33070 timesteps
and is capable of capturing the low-energy yield.

\subsection{Arbitrary two-color pulses}
Two-color laser pulses are frequently used in experiments because they allow for additional control of the electron dynamics. For instance, the net current in a plasma after the interaction with a two-color laser pulse can be optimized, which is important for the plasma-induced generation of terahertz radiation.  Two-color setups were simulated already with previous versions of \QPROP~in Refs.~\cite{Li_PhysRevLett_2017, Han_PhysRevLett_2017, Han_PhysRevLett_2018, Song_PhysRevLett_2018, Tulsky_PhysRevA_2018_THz, Tulsky_PhysRevA_2018_POP, Ge_PhysRevLett_2019, Han_PhysRevA_2019, Kruger_ApplSci_2019, Han_PhysRevLett_2019}. Here, we include illustrative examples that can be easily customized. The corresponding folders are named {\tt vortex} and {\tt lissajous}\footnote{As a test, we additionally reproduced the single-color example {\tt attoclock} within the two-color example folder and noticed no difference in the speed of the calculations.}. The laser pulse with vector potential ${\bf A}(t) = (A_x(t),A_y(t),0)$ now reads 
\begin{eqnarray}\label{AxAy}
A_x(t) &=& \frac{E_{1x}}{\omega_1} \sin^2\left(\frac{\omega_1 t}{2 n_{c1}}\right) \sin(\omega_1 t+\phi_{1x}) \\
&&  + \frac{E_{2x}}{\omega_2}\sin^2\left(\frac{\omega_2 (t-\tau)}{2 n_{c2}}\right)\sin(\omega_2 (t-\tau) +\phi_{2x}),\nonumber \\
A_y(t) &=& \frac{E_{1y}}{\omega_1} \sin^2\left(\frac{\omega_1 t}{2 n_{c1}}\right) \sin(\omega_1 t +\phi_{1y}) \nonumber \\
&&  + \frac{E_{2y}}{\omega_2}\sin^2\left(\frac{\omega_2 (t-\tau)}{2 n_{c2}}\right)\sin(\omega_2 (t-\tau) +\phi_{2y}). \nonumber
\end{eqnarray}
The pulse parameters are specified in the {\tt
  propagate.param} file. The matching of the parameter names to the
variables appearing in (\ref{AxAy}) is shown in Table~\ref{table2}. 
Although common wavelengths in the
majority of modern two-color experiments are about 800~nm, we
take for convenience $\lambda\lesssim 400$~nm in the examples here in order to keep the calculation times short.

\begin{table}[!h]
\caption{Matching of the parameter names in the {\tt propagate.param} file in the {\tt vortex} and {\tt lissajous} examples to the variables in (\ref{AxAy}).} \label{table2}
\begin{tabular}{|l|c|c|c|c|}
\hline 
\multirow{2}{*}{Parameter}           & \multirow{2}{*}{Eq.~(\ref{AxAy})} & \multirow{2}{*}{Domain}  &\multicolumn{2}{c|}{       Default in} \\ \cline{4-5}
                                     &                                  &                          & {\tt vortex}           & {\tt lissajous}       \\ \hline
{\tt omega-1}                     & $\omega_1$                       & $\mathbb {R}_{+}$        & $0.114$                   & $0.114$                   \\ \hline
{\tt omega-2}                     & $\omega_2$                       & $\mathbb {R}_{+}$        & $0.190$                   & $0.228$                   \\ \hline
{\tt max-electric-field-1-x}      &  $E_{1x}$                        &  $\mathbb {R}_{0+}$      &     $0.0267$             &       $0.0534$             \\ \hline
{\tt max-electric-field-2-x}      &  $E_{2x}$                        &  $\mathbb {R}_{0+}$      &         $0.0267$             &     $0.0$              \\ \hline
{\tt max-electric-field-1-y}      &  $E_{1y}$                        &  $\mathbb {R}_{0+}$      &         $0.0267$            &    $0.0$                \\ \hline
{\tt max-electric-field-2-y}      &  $E_{2y}$                        &  $\mathbb {R}_{0+}$      &        $0.0267$          &      $0.0534$               \\ \hline
{\tt num-cycles-1}                & $n_{c1}$                         & $\mathbb {R}_{0+}$       & $6.0$                   & $8.0$                     \\ \hline
{\tt num-cycles-2}                & $n_{c2}$                         & $\mathbb {R}_{0+}$       & $10.0$                  & $16.0$                    \\ \hline
{\tt phase-1-x}                   &  $\phi_{1x}$                  &  $\mathbb {R}$           &    $0.0$                  &    $0.0$                  \\ \hline
{\tt phase-2-x}                   &  $\phi_{2x}$                  &  $\mathbb {R}$           &    $0.0$                  &    $0.0$                  \\ \hline
{\tt phase-1-y}                   &  $\phi_{1y}$                  &  $\mathbb {R}$           &    $0.5\pi$               &    $0.0$                  \\ \hline
{\tt phase-2-y}                   &  $\phi_{2y}$                  &  $\mathbb {R}$           &    $1.5\pi$               &    $0.0$                  \\ \hline
{\tt tau-delay}                   & $\tau$                           & $\mathbb {R}_{0+}$       & $331.0$               & $0.0$                     \\ \hline
\end{tabular}
\end{table}

\subsection*{Vortex}

{\bf Time:} 25 min + (7 min $\times$ 4).

{\bf RAM:} 206 MB $\times$ 4.

\begin{figure}
\begin{center}
\includegraphics[scale=0.12]{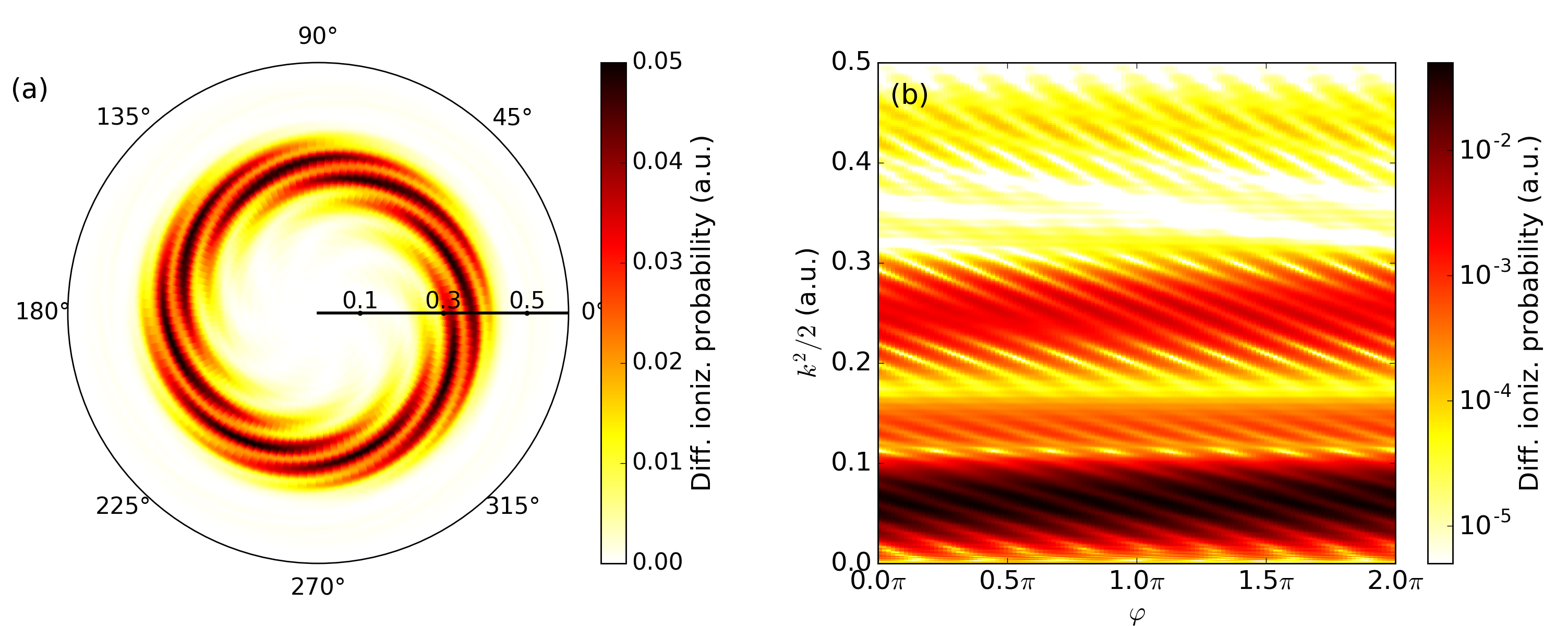}\\
\end{center}
\caption{(a) PES for hydrogen in by  $\tau=8$\,fs delayed, circularly counter-rotating laser pulses with $\lambda_1=400$\,nm and $\lambda_2 = 3 \lambda_1/5 = 240$\,nm, $I=5 \cdot 10^{13}$ Wcm$^{-2}$ each. Ticks on the horizontal black line indicate radial momenta. (b) The same  but in energy-angle coordinates and logarithmic scale.} \label{fig4}
\end{figure}

In the example located in the {\tt vortex} folder, the laser field
consists of two sequential, counter-rotating circularly polarized
pulses with $\omega_1 : \omega_2 = 3:5$ ($\lambda_1 = 400$~nm,
$\lambda_2 = 240$~nm), equal intensities $I = 5 \cdot
10^{13}$\,Wcm$^{-2}$ and equal total time duration $\tau =
8$\,fs. Note that all field component amplitudes $E_{ij}$ have to be
non-negative, and the opposite direction of the $\omega_2$ field is
achieved by adding $\pi$ to the phase $\phi_{2y}$ in the $y$
component. For simplicity, the hydrogen atom with the ionization
potential $\IP=0.5$ is used as a target. The ionization potential may
be overcome by five $\omega_1=0.114$ photons or by three
$\omega_2=0.190$ photons. Thus, photoelectrons with energy $k^2/2$ are
represented by a superposition of wavefunctions with
$Y_{55}(\theta,\varphi)\propto e^{5\imag\varphi}$ and
$Y_{3-3}(\theta,\varphi)\propto e^{-3\imag\varphi}$ angular
dependencies. Additionally, the delay $\tau$ between the pulses
creates a phase factor $e^{-\imag k^2 \tau/2}$ so that the
photoelectron yield behaves as $k\vert a({\bf k})\vert^2 \propto A + B
\cos(8\varphi-k^2 \tau/2+\alpha)$, resulting in an eight-fold
vortex-like structure (see Fig.~\ref{fig4}(a)) with a torsion defined by
$\tau$ (see the tilt in Fig.~\ref{fig4}(b)). Due to the finite duration of the
pulses, frequencies with ratios other than $3:5$ may contribute so that the total number of nodes differs for energies away from the
maximum yield. Such kind of spectra were
observed in Refs.~\cite{Pengel_PhysRevLett_2017,
  Kerbstadt_OptExpress_2017,Kerbstadt_NatureComm_2019,Xiao_PhysRevLett_2019} in which one may
also find more detailed explanations of the underlying physics.

\subsection*{Lissajous}

{\bf Time:} 14 min + (4 min $\times$ 4).

{\bf RAM:} 206 MB $\times$ 4.

\begin{figure}
\begin{center}
\includegraphics[scale=0.12]{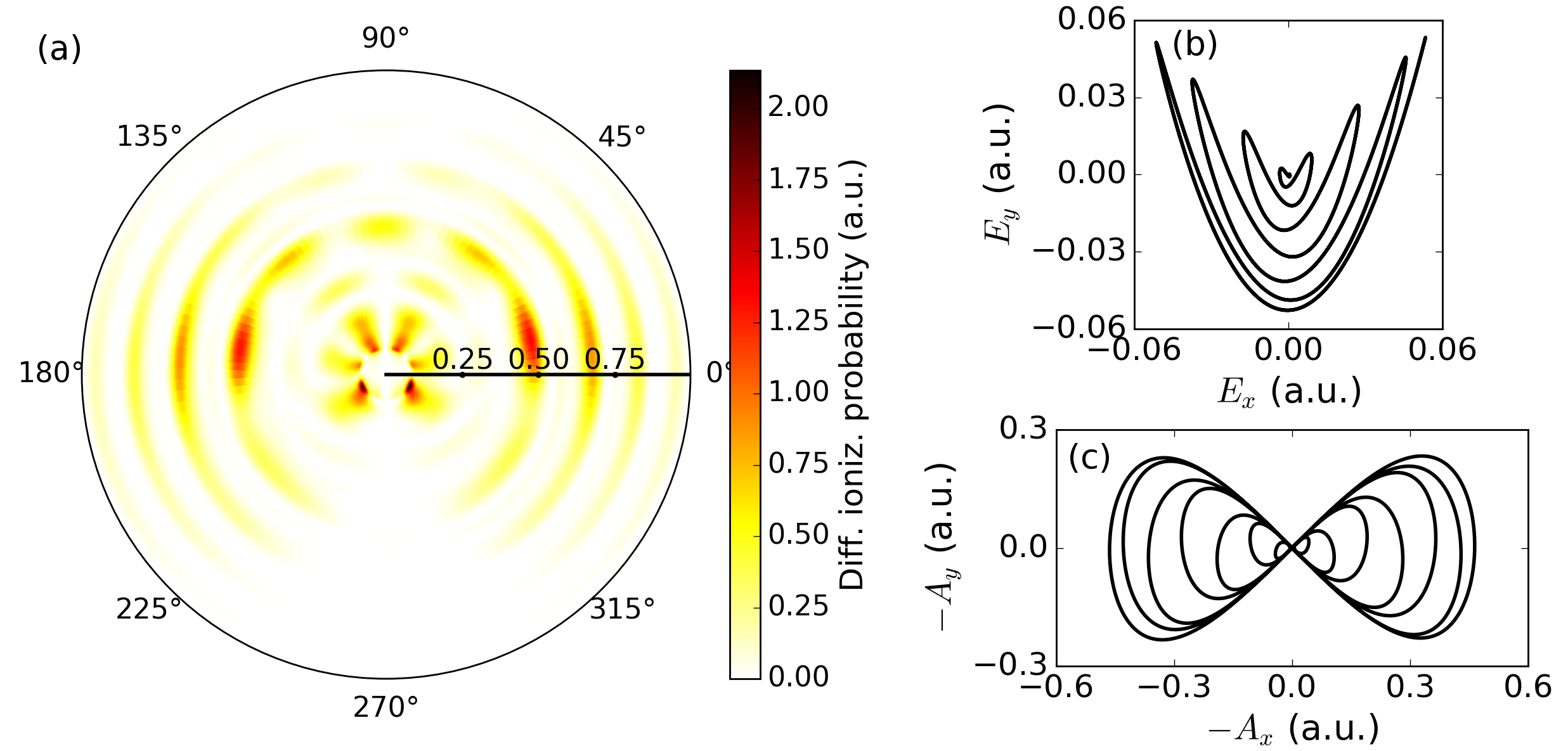}\\
\end{center}
\caption{(a) PES for hydrogen in a two-color laser pulse with orthogonal components: $x$ component with $\lambda_1=400$\,nm, $y$ component with $\lambda_2=200$\,nm, $I_1 = I_2 = 10^{14}$ Wcm$^{-2}$ and common pulse duration of $10.7$\,fs. Ticks on the horizontal black line indicate radial momenta. (b) Electric laser field ${\bf E}(t)$ and (c) its (negative) vector potential $-{\bf A}(t)$ in the polarization plane.} \label{fig5}
\end{figure}

The example in the {\tt lissajous} folder refers to another type of
laser pulses that has been used in recent theoretical and experimental studies, namely  two overlapping, orthogonally linearly polarized pulses with frequencies of ratio $n:m$ as, e.g., in Refs.~\cite{Li_PhysRevLett_2017, Han_PhysRevLett_2017,
  Han_PhysRevA_2019}. In our example, we take $\lambda_1 = 400$~nm and
$\lambda_2 = 200$~nm fields of equal intensity $I = 10^{14}$
Wcm$^{-2}$ and equal time duration $10.7$\,fs. The laser field vectors ${\bf E}(t)$ (or ${\bf A}(t)$) form Lissajous curves in the polarization plane
(see Fig.~\ref{fig5}(b,c)). Such pulse geometries allow to separate effects due to scattering off the parent ion from intracycle interferences.

\subsection*{Two-color, colinear polarization}
Two both along the $z$-axis colinearly polarized laser pulses can be simulated using {\tt qprop-dim} $=34$, which allows to use the wavefunction expansion in
spherical harmonics with fixed magnetic quantum number $m_0$
(\ref{psi2}). 
In that case the $x$ components are automatically taken as $z$
components in the code.

Such two-color, colineraly polarized pulses are, for instance, common in
experiments concerned with plasma-induced terahertz radiation \cite{Cook_JOptLett_2000} where they act as ionizing fields that prepare the initial plasma \cite{Kim_OptExpress_2007}. Another example is the addition of a weak $2\omega$ field to study the dependence of PES on the delay between strong $\omega$ and weak $2\omega$ field. Besides the additional control knob to steer the electron dynamics, such ``phase-of-the-phase'' \cite{Skruszewicz_PhysRevLett_2015, Almajid_JPhysB_2017, Tan_OptQuantEl_2018, Tulsky_PhysRevA_2018_POP,Zheng_PhysRevA_2015, Azoury2017,Porat2018, Song_PhysRevLett_2018, Eicke_PhysRevA_2019}   studies allow to reveal the coherently produced contributions to the PES that may otherwise be buried under incoherent, e.g., thermal or scattered, electrons.

\subsection{High-order harmonic generation}

{\bf Time:} 7 min.

Besides PES, strong-field physicists are often interested in the
radiation emitted by laser-illuminated targets. It is well known that
multiples of the incident laser frequency are emitted
\cite{Corkum_PhysRevLett.71.1994,Lewen_PhysRevA.49.2117} due to
so-called high-harmonic generation (HHG). Computationally, the calculation
of high-harmonic spectra is much simpler than the simulation of PES,
at least as long as a semi-classical description and the single-atom
response are sufficient. In that case only the dipole acceleration
needs to be calculated, and the PES-related t-SURFF part can be omitted. 

\begin{figure}
\begin{center}
\includegraphics[scale=0.12]{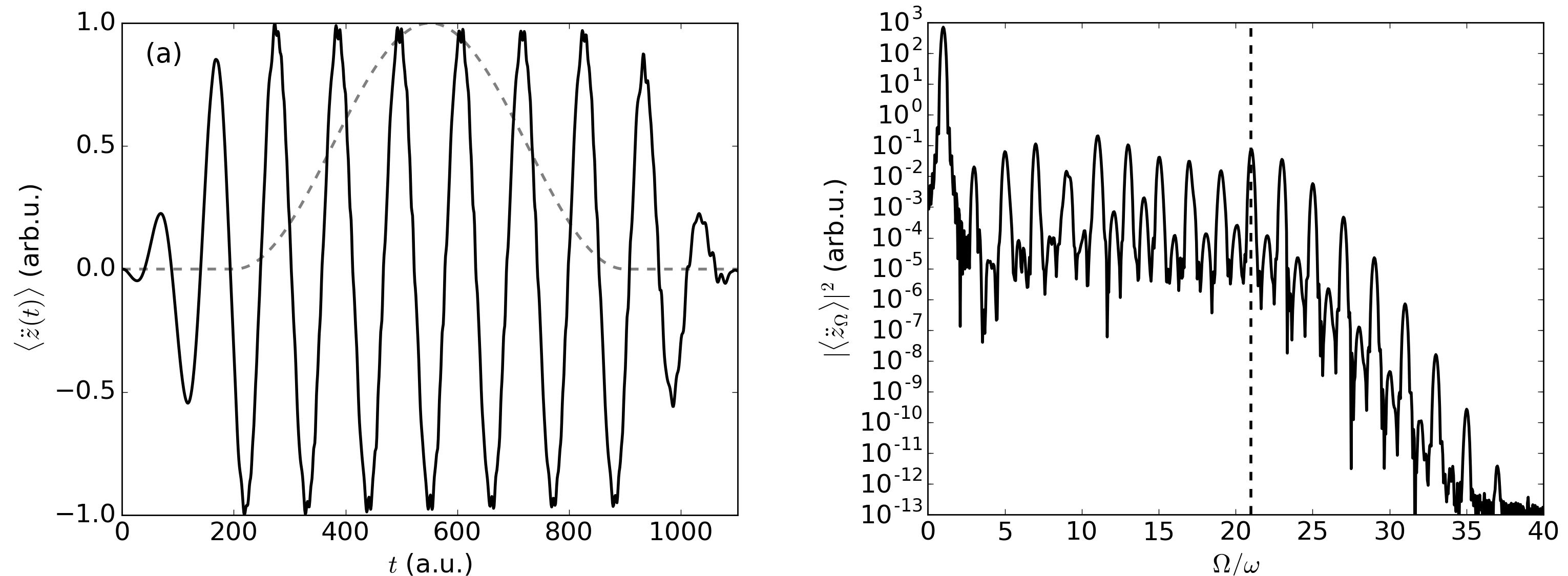}
\end{center}
\caption{(a) Dipole acceleration $\langle \ddot{d}(t) \rangle$ (solid) for the 1s electron of atomic hydrogen subject to a laser pulse (see main text for the laser parameters). The dipole acceleration is normalized to its maximum. The window function $W(t)$ is also shown (dashed). (b) Corresponding HHG spectrum. The cutoff at $\Omega = 21 \omega$ is indicated by a vertical line.} \label{fig6}
\end{figure}

We include an example for the calculation of a HHG spectrum in the
folder {\tt hhg}. There, a linearly polarized pulse with an
$n_c-4=6$-cycle flat-top part  and 2-cycle $\sin^2$-shape up and down rampings impinges on a hydrogen atom. Wavelength and intensity are $\lambda = 800$\,nm and $I = 10^{14}$ Wcm$^{-2}$, respectively.  
Figure~\ref{fig6} shows the expectation value of the
acceleration as a function of time, calculated as
\begin{equation}
\langle \ddot{z}(t) \rangle  = -\left\langle \Psi(t) \left\vert \frac{\partial U(r)}{\partial z}  \right\vert \Psi(t) \right\rangle - E_z(t),
\end{equation}
and the HHG spectrum  $\left\vert \langle \ddot{d}_\Omega \rangle \right\vert^2 $ with
\begin{equation}
 \langle \ddot{d}_\Omega \rangle = - \int_0^{\infty} W(t)\langle \ddot{z}(t) \rangle  e^{-\imag\Omega t} dt .
\end{equation}
The window function $W(t)$ was chosen $\sin^{2}$ in the flat-top region of the laser pulse and zero outside. That allows to obtain very pronounced HHG peaks with a high signal-to-noise ratio.  The HHG spectrum consists of sharp peaks at odd multiples of the incident laser frequency, forming the so-called plateau up to the cutoff $\Omega = \IP+3.17 \UP $ \cite{Krause_PhysRevLett_1992,Lewen_PhysRevA.49.2117} and a fast decrease thereafter.

For a laser pulse polarized in the $xy$ plane, the HHG spectrum is calculated according 
\begin{equation}\label{HHG_xy}
\left\vert \langle \ddot{\bf d}_\Omega \rangle \right\vert^2 = \left\vert \langle \ddot{x}_\Omega \rangle \right\vert^2 + \left\vert \langle \ddot{y}_\Omega \rangle \right\vert^2
\end{equation}
where first and second term are calculated separately from the Fourier-transforms of the real and imaginary parts of 
\begin{equation}
\langle \ddot{x}(t)+\imag~\ddot{y}(t) \rangle = -\left\langle \Psi(t) \left\vert \frac{\partial U(r)}{\partial x}+\imag \frac{\partial U(r)}{\partial y}  \right\vert \Psi(t) \right\rangle - (E_x(t) + \imag E_y(t)).
\end{equation}
The ellipticity of a certain harmonic can be obtained from the phase difference  of the Fourier transforms at the respective $\Omega$ value. 

The reader may, for instance, reconsider the {\tt lissajous} example,
switch the {\tt generate-hhg-data} trigger from 0 to 1, the {\tt
  generate-tsurff-data} trigger from 1 to 0 and set {\tt pot-cutoff}
larger than the grid size to avoid a potential truncation (required for t-SURFF but not for HHG). In such an orthogonal, two-color $\omega$-$2\omega$ setup
   the dipole acceleration expectation value in $y$ direction  consists of
even harmonics of the fundamental frequency $\omega$ (in $x$ direction). Therefore, the sum (\ref{HHG_xy}) yields a HHG spectrum with both odd and even multiples of $\omega$. For a more detailed study and applications
see, e.g., Refs.~\cite{Shafir_Nature_2009, Niikura_PhysRevLett_2010,
  Brugnera_PhysRevLett_2011, Niikura_PhysRevLett_2011,
  Murakami_PhysRevA_2013}.

\section{{\bf Plotting guide}} \label{sec:plotguide}

For the user's convenience, scripts written in Python that were used to
visualize the \QPROP~3.0 generated data are added to the package. They
are located in {\tt scr/plots}.

Photoelectron distributions can be plotted using {\tt plot\_pes.py}. Leave the desired filename uncommented in the upper section of it, choose the number of angles and the {\tt polarization} as {\tt ='xz'} for linear or {\tt ='xy'} otherwise. 

Figure~\ref{fig1}(a) was produced with {\tt polar\_canvas=1}, Fig.~\ref{fig1}(c) with {\tt plot\_type='1D'}, Fig.~\ref{fig2}(a) was based on the same  script with {\tt polar\_canvas=0} and {\tt cartesian=1}. Figure~\ref{fig4}(b) shows a spectrum plotted with respect to energy, hence {\tt wrt\_energy=1}, and a canvas format {\tt polar\_canvas=0}, {\tt cartesian=0}. Laser field plots in Fig.~\ref{fig5}(b,c) are generated with the script {\tt plot\_laser.py}. Plots in Fig.~\ref{fig6} are created with a slight modification of the script {\tt plot\_hhg.py}.

\section*{Acknowledgment}

This work was supported by the project BA 2190/10 of the German Science Foundation (DFG).





\section*{References}

\bibliographystyle{elsarticle-num}
\bibliography{biblio}







\end{document}